\documentclass[twocolumn,aps,prl,floats,floatfix,superscriptaddress]{revtex4-2}
\usepackage{amsmath}
\usepackage{graphicx}
\usepackage{dcolumn}
\usepackage[english]{babel}

\usepackage{color}
\usepackage{float}

\usepackage{amssymb}

\begin{document}

\title{Localization versus hybridization of $f$ states in actinide and lanthanide dioxides probed in core-level photoemission spectra}

\author{Sergei M. Butorin}
\affiliation{Condensed Matter Physics of Energy Materials, X-ray Photon Science, Department of Physics and Astronomy, Uppsala University, P.O. Box 516, SE-751 20 Uppsala, Sweden}
\email{sergei.butorin@physics.uu.se}



\begin{abstract}
The degrees of the localization and hybridization of the valence $f$ states/covalency of the chemical bonding in actinide and lanthanide dioxides were investigated using the atomic, crystal-field multiplet and Anderson impurity model (AIM) approaches to calculate actinide $5d$ and lanthanide $3d$ x-ray photoemission spectra (XPS). The actinide $5d$ XPS can be largely described within atomic, crystal-field multiplet theory due to an extended multiplet structure as a result of the strong interaction of $5f$ electrons with a $5d$ core hole. The multiplet structure was found to be quite sensitive to the oxidation state of actinides. In turn, the lanthanide $3d$ XPS description requires the AIM-type of calculations due to significant $4f-$O $2p$ hybridization effects. As a result derived from the analysis of the XPS spectra, an increase in the $f$-shell occupancy in the ground state due to the $f-$O $2p$ hybridization and covalency of the chemical bonding appears to be higher in lanthanide dioxides as compared to actinide dioxides.
\end{abstract}


\maketitle

\section{Introduction}

The degree of the localization is an important aspect when it comes to a discussion of the properties of $f$ states of actinides and lanthanides. In general, the high degree of the localization is expected for the $4f$ states of lanthanides while the $5f$ states of actinides often reveal the so-called dual (localized versus itinerant) character depending on the system. One of the efficient means to evaluate the degree of the $f$-localization is to employ x-ray spectroscopy for the electronic structure investigation. For insulating compounds, the multiplet approach is extensively used for the description of the electronic structure and x-ray spectroscopic data.

X-ray absorption and photoemission spectroscopies (XAS and XPS, respectively) at core levels of actinides and lanthanides help to determine the oxidation state of these elements in various compounds. They also provide signatures of the hybridization of the $f$ states with ligand states in terms of so-called actinide $5f-$ligand $p$ or lanthanide $4f-$ligand $p$ charge-transfer satellites, especially in the core-level XPS spectra. The analysis of such spectra in terms of calculations to reproduce the experimental data allows for the estimation of the $f$-count (number of $f$ electrons $n_f$) in compounds in question.

For actinides, the XPS measurements are most often performed at the $4f$ levels (see e.g. reviews in Refs. \cite{Ilton,Caciuffo}) because the corresponding transitions have a large cross-section. However, most of the multiplet structure is hidden by the core-hole lifetime broadening of the spectra (see e.g. \cite{Butorin_UC,Butorin_Am}). Therefore, the actinide oxidation states are determined by the chemical shifts of the $4f$ XPS lines, often through the fitting procedure of multiple components, or by an appearance of the charge-transfer satellites at specific binding energies in the case of higher actinide valencies.

On the other hand, the XPS transitions at the actinide $5d$ levels are represented by the extended multiplet structure with a significant spread in energy due to a strong interaction between the $5d$ core hole and $5f$ electrons. This was already pointed out for uranium oxides \cite{Beatham,Ilton2}. Therefore, the characteristic profiles of the multiplet structure in the actinide $5d$ XPS spectra can help to distinguish between different oxidation states. Although, the actinide $5d$ XPS data are rarely measured in detail, a number of spectra have been already reported for various actinide oxides \cite{Beatham,Ilton2,Veal,Allen,Teterin_PuO2,Teterin_NpO2,Teterin_AmO2,Putkov}. The goal of this paper is to better understand the utility of such data by calculating the $5d$ XPS spectra for Th, U, Np, Pu, Am and Cm dioxides using the atomic, crystal-field multiplet theory and Anderson impurity model (AIM) \cite{Anderson}.

Although, the high degree of the $4f$ localization is expected in insulating compounds of lanthanides, there are a few dioxides, such as CeO$_2$, PrO$_2$ and TbO$_2$, where the significant $4f$ hybridization with O $2p$ states was established. The hybridization leads to an appearance of the lanthanide $4f-$O $2p$ charge-transfer satellites in the $3d$ XPS spectra of these dioxides. For CeO$_2$, the analysis of the Ce $3d$ XPS spectrum was done by including the full atomic multiplet effects within AIM \cite{Kotani,Nakazawa}. Based on that, the Ce $4f$ occupancy in CeO$_2$ was estimated to be $n_f$=0.56 electrons which clearly deviates from the pure ionic picture of the Ce $4f^0$ configuration in the ground state. This estimation was supported by the corresponding analysis of the Ce $3d$-$4f$ resonant inelastic x-ray scattering (RIXS) spectra of CeO$_2$ \cite{Butorin_CeO2} which probe the electronic structure of the ground state due to an absence of a core hole in the final state of the spectroscopic process. Recent calculations of the Ce $3d$ XPS spectrum of CeO$_2$ using a combination of density functional theory and dynamic mean-field theory (DFT+DMFT) \cite{Kolorenc} suggested the $n_f$ value of 0.44 electrons. For PrO$_2$ and TbO$_2$, the analysis of the $3d$ XPS spectra was done within AIM \cite{Kotani,Ikeda} but without including the full multiplet effects. Although, the results in Refs. \cite{Kotani,Ikeda} already indicated a significant covalency of the Pr--O and Tb--O bonds in PrO$_2$ and TbO$_2$, taking into account the multiplet structure in the AIM calculation is desirable in order to reproduce both $3d_{5/2}$ and $3d_{3/2}$ XPS manifolds and provide more reliable quantitative estimates.


\section{Computational details}
For actinide oxides, as the first step, the crystal-field multiplet approach was used in the $5d$ XPS calculations which included the $5f$ and core $5d$ states on a single An ion in cubic symmetry. The total Hamiltonian of the system included the the Coulomb, exchange and spin-orbit interactions for a free actinide ion as well as interactions leading to the crystal-field splittings of the $5f$ shell, as described in Refs. \cite{Butorin_UO2,Butorin_3d4fRIXS}. For the $5d$ XPS, the ground and final states of the spectroscopic process are represented by the $5f^{n}$ and $5d^{9}5f^{n}\varepsilon{l}$ electronic configurations, respectively ($\varepsilon{l}$ corresponds to an electron in the continuum). The interactions between $5f$ electrons and between a core $5d$ hole and $5f$ electrons are described in terms of Slater integrals $F^{2,4,6}(5f,5f)$, $F^{2,4}(5d,5f)$ and $G^{1,3,5}(5d,5f)$, while the spin-orbit interactions for the $5f$ and core $5d$ states are described with coupling constants $\zeta(5f)$ and $\zeta(5d)$, respectively. The electrostatic and spin-orbit interactions of the continuum electron are taken equal to zero. The cubic crystal-field potential and the values of the crystal-field $5f$ splittings are characterized by Wybourne's crystal-field parameters $B^{4}_{0}$ and $B^{6}_{0}$.

The actinide $5d$ XPS spectra were calculated using the following equation
\begin{eqnarray}
I_{XPS}(E) = \sum_{f} | \langle f | a_c | g \rangle |^{2} \frac{\Gamma/\pi}{(E_{f}-E_{g}-E)^{2}+\Gamma^{2}},
\end{eqnarray}
where $| g \rangle$ and $| f \rangle$ are the ground and XPS final states of the spectroscopic process with energies $E_{g}$ and $E_{f}$, respectively. $E$ is the binding energy, and $a_c$ is the annihilation operator of a core electron and $\Gamma$ is a lifetime broadening of the XPS final state in terms of half width at half maximum (HWHM).

As the first step, the constant $\Gamma$ value throughout the $5d$ threshold was used in the calculations which leads to the uniform broadening of the calculated spectra. However, in reality, $\Gamma$ significantly varies for each excited state of the $5d^{9}5f^{n}$ ($n\geq{1}$) configuration as a result of Coster-Kronig $\langle5d^{9}5f^{n}|1/r|5d^{10}5f^{n-1}6(s, p)^{-1}\varepsilon{l}\rangle$ and super Coster-Kronig $\langle5d^{9}5f^{n}|1/r|5d^{10}5f^{n-2}\varepsilon{l}\rangle$ decays. To take that into account, the XPS calculations were performed using the approach described by Ogasawara \textit{et al.} \cite{Ogasawara}. As a result, $\Gamma$ is determined as
\begin{eqnarray}
\Gamma = \pi\sum_{A} | \langle A | H_A | f \rangle |^{2} \delta(E_A-E_f),
\end{eqnarray}
where $| A \rangle$ is an Auger decayed state with energy $E_A$. $| f \rangle$ is coupled to $| A \rangle$ via the configuration interaction represented by $H_A$.

To take into account the hybridization effects of the $f$ states with O $2p$ states in the XPS calculations for actinide and lanthanide oxides, AIM was used. The AIM Hamiltonian of a system, including the multiplet effects, can be written as
\begin{eqnarray}
H&=&\varepsilon_{f}\sum_{\gamma} a^{\dag}_{5f}(\gamma)a_{5f}(\gamma) \nonumber  \\
       &+&
\varepsilon_{d}\sum_{\mu} a^{\dag}_{d}(\mu)a_{d}(\mu) \nonumber  \\
       &+&
\sum_{\sigma,\gamma} \varepsilon_{\upsilon}(\sigma)a^{\dag}_{\upsilon}(\sigma,\gamma)a_{\upsilon}(\sigma,\gamma) \nonumber  \\
       &+& U_{ff}\sum_{\gamma>\gamma^{\prime}}a^{\dag}_{f}(\gamma)a_{f}(\gamma)a^{\dag}_{f}(\gamma^{\prime})a_{f}(\gamma^{\prime})  \nonumber\\
       &-&
       U_{fc}\sum_{\gamma,\mu}a^{\dag}_{f}(\gamma)a_{f}(\gamma)a^{\dag}_{d}(\mu)a_{d}(\mu) \nonumber\\
       &+&
       \frac{V}{\sqrt{N}}\sum_{\sigma,\gamma} [(a^{\dag}_{\upsilon}(\sigma,\gamma)a_{f}(\gamma) + a^{\dag}_{f}(\gamma)a_{\upsilon}(\sigma,\gamma)] \nonumber\\
       &+&
       H_{multiplet},
\end{eqnarray}
where $\varepsilon_{f}$, $\varepsilon_{d}$, and $\varepsilon_{\upsilon}$ are one-electron energies of the $f$, core $d$ and valence band levels, respectively, and $a^{\dag}_{f}(\gamma)$, $a^{\dag}_{d}(\mu)$, $a^{\dag}_{\upsilon}(\sigma,\gamma)$ are electron creation operators at these levels with combined indexes $\gamma$ and $\mu$ to represent the spin and orbital states of the $f$, $d$ and valence-band electrons, $\sigma$ is the index of the $N$ discrete energy levels in the O $2p$ band (bath states). $U_{fc}$ is the core $d$ hole potential acting on the $f$ electrons. $V$ is the hybridization strength (or hopping term) between the $f$ states and states of the O $2p$ band. The $\varepsilon_{\upsilon}(\sigma)$ is represented by the $N$ discrete levels/bath states in the form
\begin{eqnarray}
\varepsilon_{\upsilon}(\sigma)=\varepsilon_{\upsilon}^0-\frac{W}{2}+\frac{W}{N}(\sigma-\frac{1}{2}),~\sigma=1,...,N,
\end{eqnarray}
where $\varepsilon_{\upsilon}^0$ and $W$ are the center and width of the O $2p$ band, respectively. In calculations for actinide and lanthanide oxides, the core $d$ level was represented by $5d$ and $3d$, respectively.

The \textit{ab-initio} values of Slater integrals $F^{k}$, $G^{k}$, $R^{k}$, spin-orbit coupling constants and matrix elements were obtained with the TT-MULTIPLETS package which combines Cowan's atomic multiplet program \cite{Cowan} (based on the Hartree-Fock method with relativistic corrections) and Butler's point-group program \cite{Butler}, which were modified by Thole \cite{Thole}, as well as the charge-transfer program written by Thole and Ogasawara.

To compare with the experimental data, it is usually necessary to uniformly shift the calculated spectra on the energy scale because it is difficult to accurately reproduce the absolute energies in this type of calculations.

\section{Results and discussion}
\subsection{Actinide $5d$ XPS}

\begin{figure}[t]
\includegraphics[width=\columnwidth]{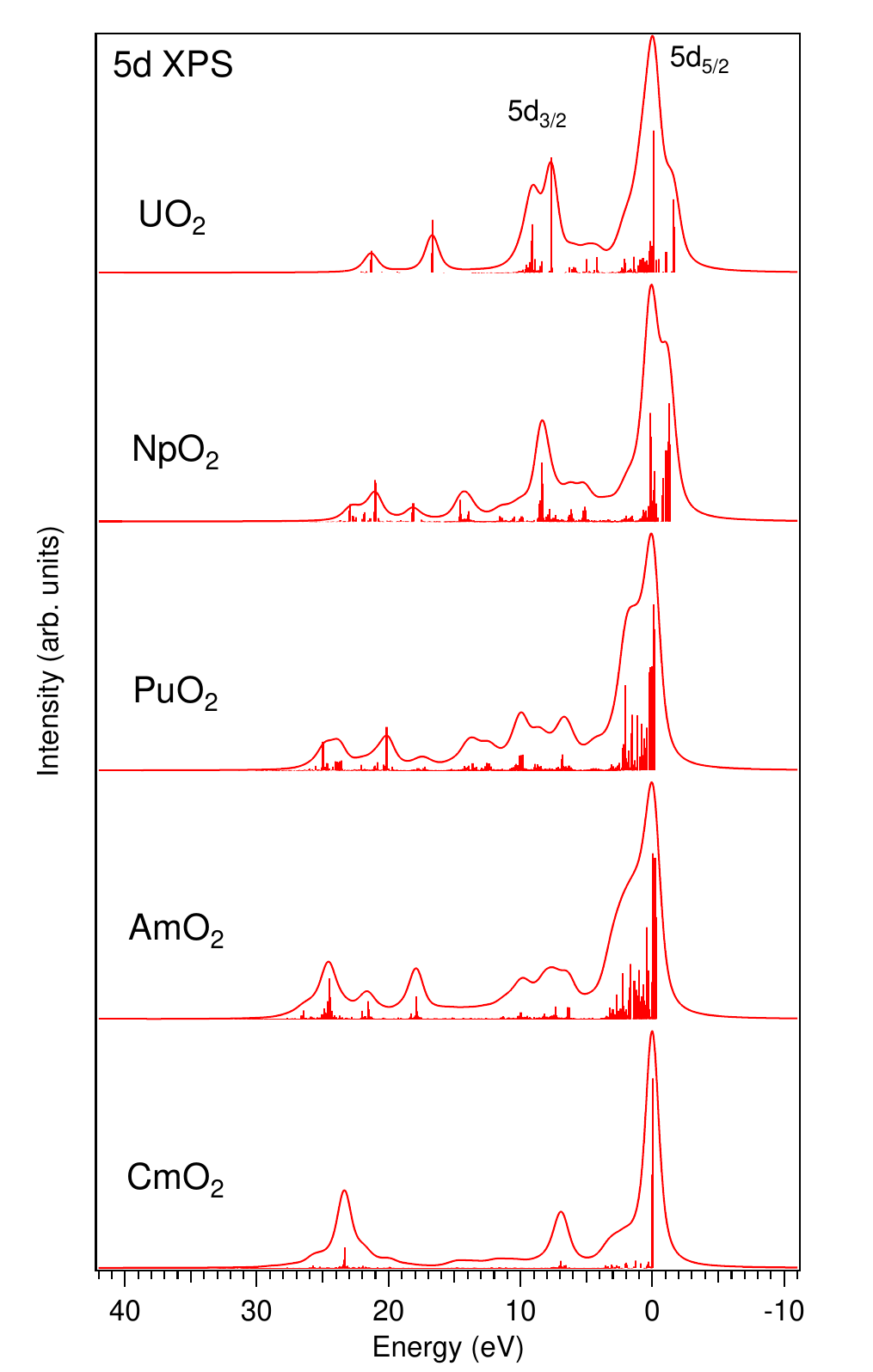}
\caption{Calculated $5d$ XPS spectra of several actinide dioxides using crystal-field multiplet theory. Vertical sticks represent multiplet poles. Solid lines are convoluted spectra. The spectra are aligned at their main maxima which are set to zero eV, for the purpose of comparison. \label{AnO2_Oh_5dXPS}}
\end{figure}

Fig.~\ref{AnO2_Oh_5dXPS} displays the $5d$ XPS spectra of several actinide dioxides calculated using crystal-field multiplet theory. The ground and final states of the spectroscopic process were described by $5f^{n}$ and $5d^{9}5f^{n}\varepsilon{l}$ ($n$=2,3,4,5,6) configurations, respectively. The \textit{ab-initio} Hartree-Fock (in the intermediate coupling approximation) values of Slater integrals $F^{2,4,6}(5f,5f)$, $F^{2,4}(5d,5f)$, $G^{1,3,5}(5d,5f)$, and spin-orbit coupling constants $\zeta(5f)$ and $\zeta(5d)$ for these configurations are listed in Tables~\ref{table1},~\ref{table2}. In the $5d$ XPS calculations, $F^{k}(5f,5f)$, $F^{k}(5d,5f)$ and $G^{k}(5d,5f)$ were reduced to 80\%, 75\% and 65\%, respectively, of their \textit{ab-initio} values to account for the intra-atomic correlation and solid state effects as it is usually done in this type of calculations. The spin-orbit interaction was kept at full strength. The amounts of reduction of Slater integrals are almost the same as those used in the $4d$ XPS calculations for lanthanide oxides \cite{Ogasawara} as well as in the calculations of the actinide $5d$ XAS spectra \cite{Ogasawara_5dXAS,Moore,Butorin_Am,Butorin_review,Butorin_nano,Butorin_AnalChem,Butorin_AmRIXS}. The values of cubic crystal field parameters $B^{4}_{0}$ and $B^{6}_{0}$ for the $5f$ shell are listed in Tables~\ref{table3} and are the same as those used for the calculated $M_{4,5}$ HERFD-XAS (high energy resolution fluorescence detection x-ray absorption spectroscopy) spectra of these actinide dioxides \cite{Butorin_3d4fRIXS}. For spectra in Fig.~\ref{AnO2_Oh_5dXPS}, the core-hole lifetime broadening was assumed to be constant for all the poles of the final state multiplet so that the spectra were broadened by the Lorentzian with $\Gamma$=0.4 eV. In addition, the experimental resolution was simulated by the Gaussian with HWHM of 0.3 eV.

\begin{table}[b]
\caption{\textit{Ab-initio} Hartree-Fock values (in units of eV) of Slater integrals and spin-orbit coupling constants in the ground-state configurations ($5f^n$) of actinide ions. In the XPS calculation, the Slater integrals were reduced to 80\% of these values.}
\begin{tabular}{lccccc}
Ion&$n$&$F^{2}(5f,5f)$&$F^{4}(5f,5f)$&$F^{6}(5f,5f)$&$\zeta(5f)$\\
\hline
U(IV)&2&9.514&6.224&4.569&0.261\\
Np(IV)&3&9.907&6.489&4.767&0.297\\
Pu(IV)&4&10.282&6.741&4.955&0.334\\
Am(IV)&5&10.642&6.982&5.136&0.373\\
Cm(IV)&6&10.990&7.216&5.310&0.414\\
Cm(III)&7&10.456&6.825&5.008&0.386\\
\end{tabular}
\label{table1}
\end{table}

\begin{table*}
\caption{\textit{Ab-initio} Hartree-Fock values (in units of eV) of Slater integrals and spin-orbit coupling constants in the final-state configurations ($5d^{9}5f^{n}\varepsilon{l}$) of actinide ions. In the XPS calculation, the $F^{k}(5f,5f)$, $F^{k}(5d,5f)$ and $G^{k}(5d,5f)$ integrals were reduced to 80\%, 75\% and 65\% of these values, respectively.}
\begin{tabular}{lcccccccccc}
Ion&$F^{2}(5f,5f)$&$F^{4}(5f,5f)$&$F^{6}(5f,5f)$&$\zeta(5f)$&$F^{2}(5d,5f)$&$F^{4}(5d,5f)$&$G^{1}(5d,5f)$&$G^{3}(5d,5f)$&$G^{5}(5d,5f)$&$\zeta(5d)$\\
\hline
U(IV)&10.228&6.744&4.971&0.299&11.138&7.215&13.233&8.214&5.873&3.237\\
Np(IV)&10.581&6.980&5.147&0.334&11.545&7.500&13.785&8.562&6.125&3.508\\
Pu(IV)&10.925&7.210&5.319&0.372&11.938&7.776&14.316&8.898&6.369&3.784\\
Am(IV)&11.250&7.426&5.480&0.411&12.310&8.035&14.816&9.214&6.597&4.075\\
Cm(IV)&11.582&7.649&5.646&0.454&12.684&8.296&15.317&9.531&6.827&4.383\\
Cm(III)&11.131&7.316&5.387&0.426&12.282&7.991&14.758&9.159&6.551&4.342\\
\end{tabular}
\label{table2}
\end{table*}

\begin{table}[b]
\caption{Values (in units of eV) of Wybourne's crystal-field parameters for the $5f$ shell used in the XPS calculations for actinide dioxides.}
\begin{tabular}{lccccc}
Parameter&U(IV)&Np(IV)&Pu(IV)&Am(IV)&Cm(IV)\\
\hline
$B^{4}_{0}$&-0.93&-0.84&-1.21&-0.84&-0.80\\
$B^{6}_{0}$&0.35&0.34&0.50&0.27&0.15\\
\end{tabular}
\label{table3}
\end{table}

The difference of the actinide $5d$ XPS spectra from the $4f$ XPS spectra is that there is no a clear distinction and separation between the $5d_{5/2}$ and $5d_{3/2}$ components due to an extended multiplet structure. While the $4f_{7/2}$ and $4f_{5/2}$ XPS lines are often similar in shape, the profile of the $5d_{3/2}$ region in the $5d$ XPS spectra differs from that of $5d_{5/2}$. An inspection of published experimental $5d$ XPS data of actinide oxides \cite{Beatham,Ilton2,Veal,Allen,Teterin_PuO2,Teterin_NpO2,Teterin_AmO2,Putkov} reveals that the calculated spectra in Fig.~\ref{AnO2_Oh_5dXPS} describe the experimental XPS structures fairly well (except for CmO$_2$) in the region from -11 to $\sim$15 eV. That is not surprising because the use of the atomic and crystal-field multiplet approach was also sufficient for the description of the actinide RIXS spectra at the $5d$ edges \cite{Butorin_Am,Butorin_review,Butorin_nano,Butorin_AnalChem,Butorin_AmRIXS,Kvashnina}.

\begin{figure}
\includegraphics[width=\columnwidth]{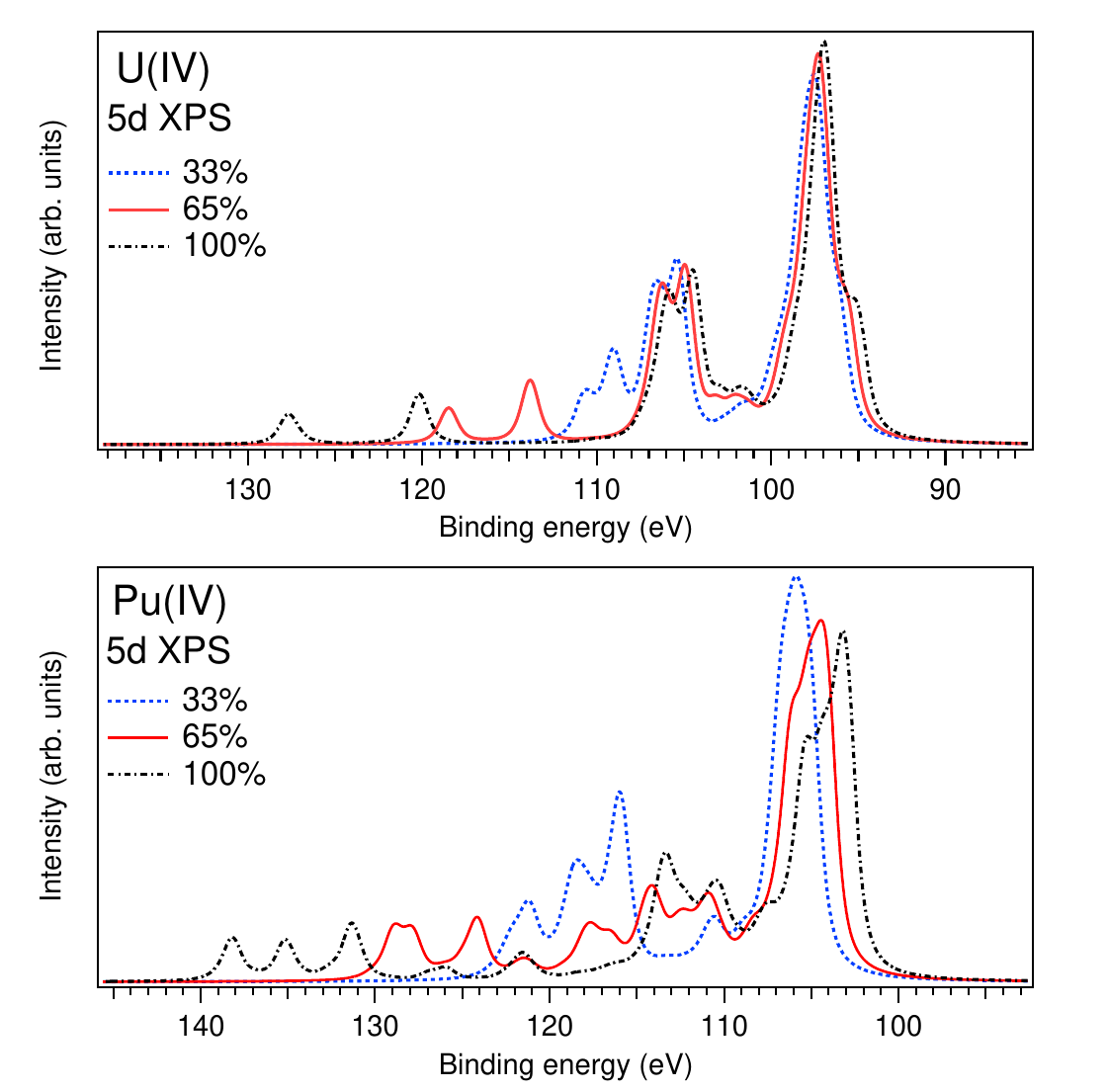}
\caption{Calculated $5d$ XPS spectra of U(IV) and Pu(IV) when $G^k$ integrals were reduced to 33\%, 65\% of their \textit{ab-initio} Hartree-Fock values and when they were kept at full strength (100\%). \label{AnO2_atomic}}
\end{figure}

However, the calculated $5d$ XPS spectra in Fig.~\ref{AnO2_Oh_5dXPS} also show sharp structures at higher energies (15-28 eV) which are not really observed in the experimental spectra. These extra structures are mainly a result of the strong $5d$-$5f$ exchange interaction characterized by the $G^k$ Slater integrals. To illustrate this, Fig.~\ref{AnO2_atomic} displays the $5d$ XPS spectra of U(IV) and Pu(IV) calculated with $G^k$ integrals reduced to 33\% and 65\% of their \textit{ab-initio} Hartree-Fock values as well as with $G^k$ integrals at their full strength (100\%). For U(IV), one can see two peaks which move away from the main spectral structures towards higher binding energies with increasing $G^k$ values. The existence and character of these peaks was already discussed by Beatham \textit{et al.} \cite{Beatham}. For Pu(IV), the profile of such extra structures appears to be more complex but their energy separation from the main spectral structures also increases for larger $G^k$ values.

Similar structures at high binding energies were found in the atomic multiplet calculations of the $4d$ XPS spectra of lanthanide oxides \cite{Ogasawara}. It was shown that these structures are significantly smeared out in the experimental $4d$ XPS spectra due to an enhanced probability for the $4d$-$4f4f(\varepsilon{l})$ super Coster-Kronig decay. To take such effects into account, the actinide $5d$ XPS spectra were calculated using the formalism described by Ogasawara \textit{et al.} \cite{Ogasawara}. Since the crystal-field effects do not lead to any significant changes in the shape of these spectra (compare Figs.~\ref{AnO2_Oh_5dXPS} and \ref{AnO2_atomic}) the atomic multiplet approach was used for simplicity. The dominant $5d$-$5f5f(\varepsilon{g})$ super Coster-Kronig decay channel was considered and Slater integrals $R^k(5f5f,5d\varepsilon{g}$) for this decay were reduced to 80\% of their original values listed in Table~\ref{table4}. The values of the $F^{k}(5f,5f)$, $F^{k}(5d,5f)$ and $G^{k}(5d,5f)$ integrals were kept the same as in the calculations of the spectra in Fig.~\ref{AnO2_Oh_5dXPS}. Varying core-hole lifetime $\Gamma$ for each multiplet state of the $5d^95f^n\varepsilon{l}$ configuration was determined using equation 2.

\begin{figure}
\includegraphics[width=\columnwidth]{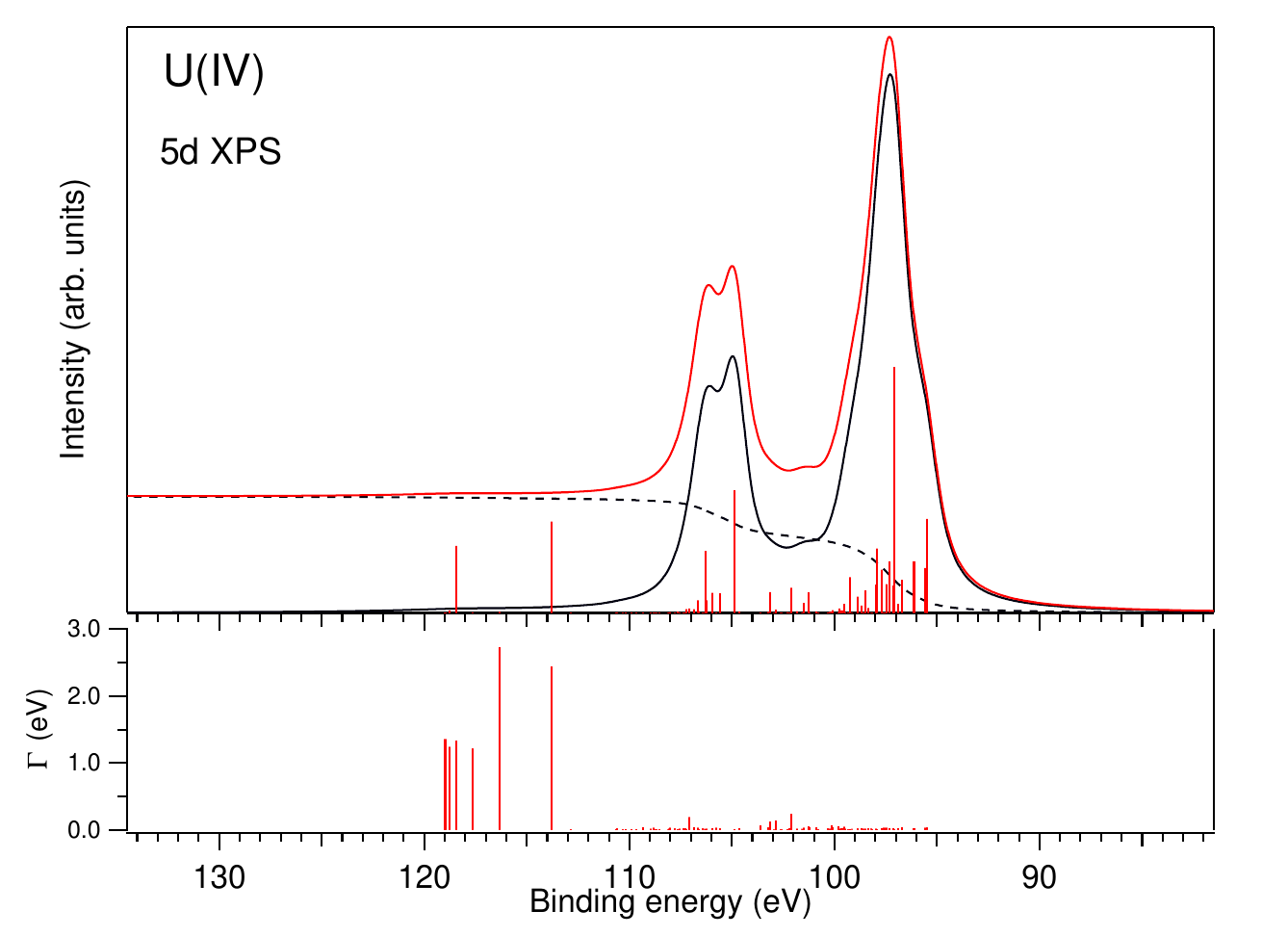}
\caption{Calculated $5d$ XPS spectrum of U(IV) by taking into account the Auger decay of states of the $5d^95f^2\varepsilon{l}$ configuration and corresponding multiplet dependent lifetime broadening $\Gamma$. The solid black curve is the convoluted spectrum, the dashed black curve represents the background \cite{Ikeda} and the solid red curve is the sum of the convoluted spectrum and background. $\Gamma$ is calculated for all multiplet states including those which have no intensity. \label{UO2_5dXPS}}
\end{figure}

\begin{figure}
\includegraphics[width=\columnwidth]{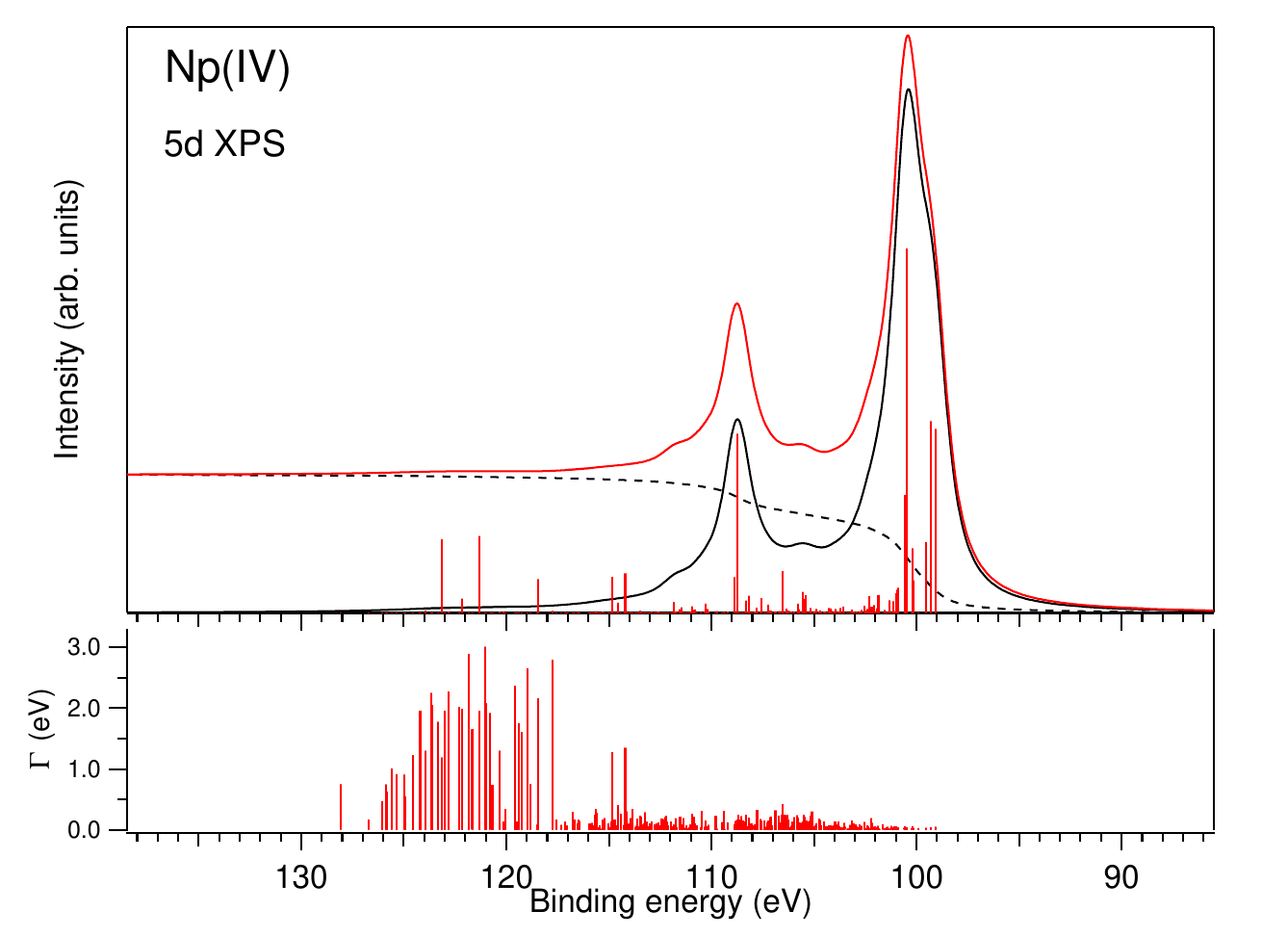}
\caption{Calculated $5d$ XPS spectrum of Np(IV) by taking into account the Auger decay of states of the $5d^95f^3\varepsilon{l}$ configuration and corresponding multiplet dependent lifetime broadening $\Gamma$. \label{NpO2_5dXPS}}
\end{figure}

\begin{figure}
\includegraphics[width=\columnwidth]{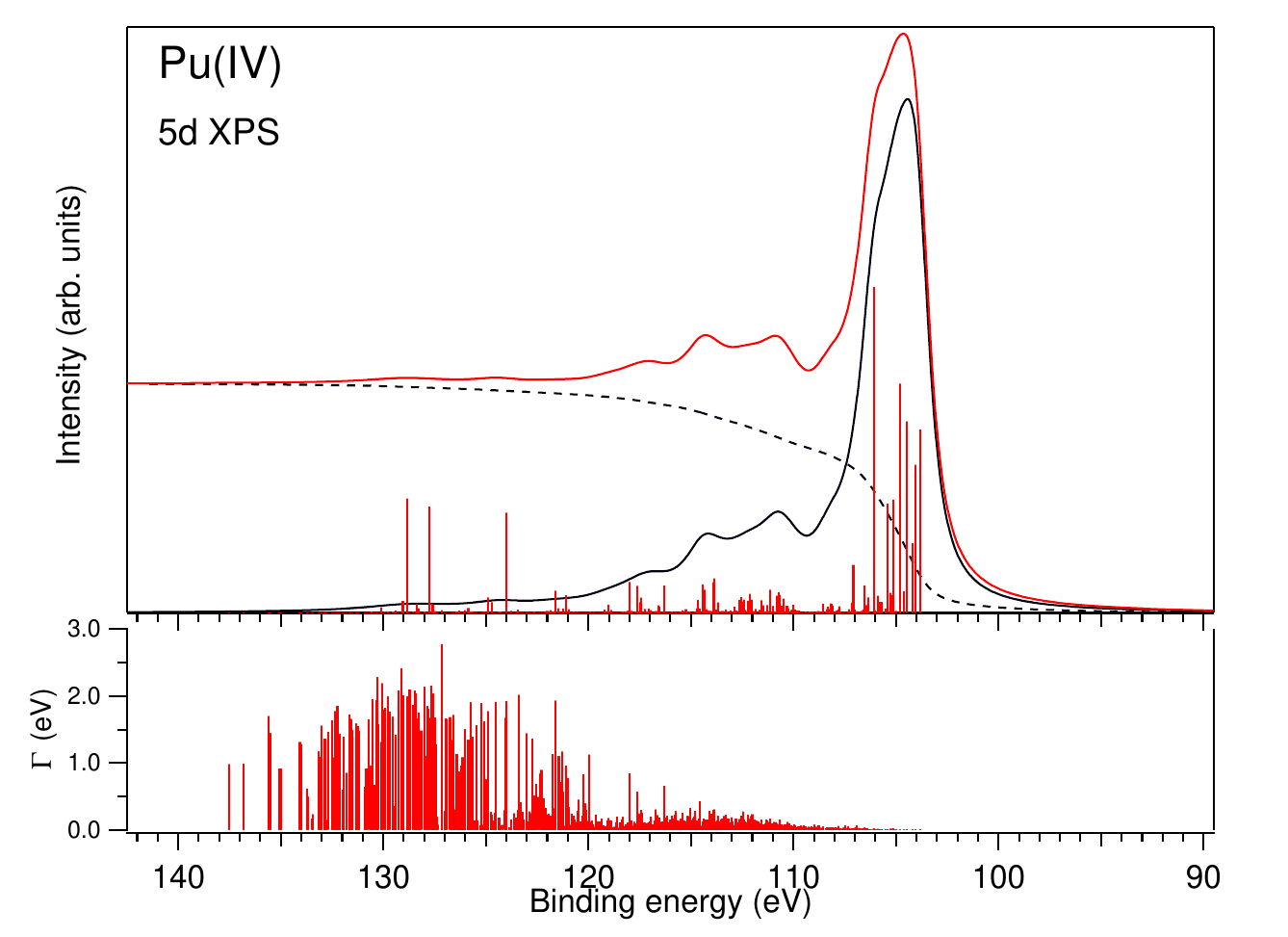}
\caption{Calculated $5d$ XPS spectrum of Pu(IV) by taking into account the Auger decay of states of the $5d^95f^4\varepsilon{l}$ configuration and corresponding multiplet dependent lifetime broadening $\Gamma$. \label{PuO2_5dXPS}}
\end{figure}

\begin{figure}
\includegraphics[width=\columnwidth]{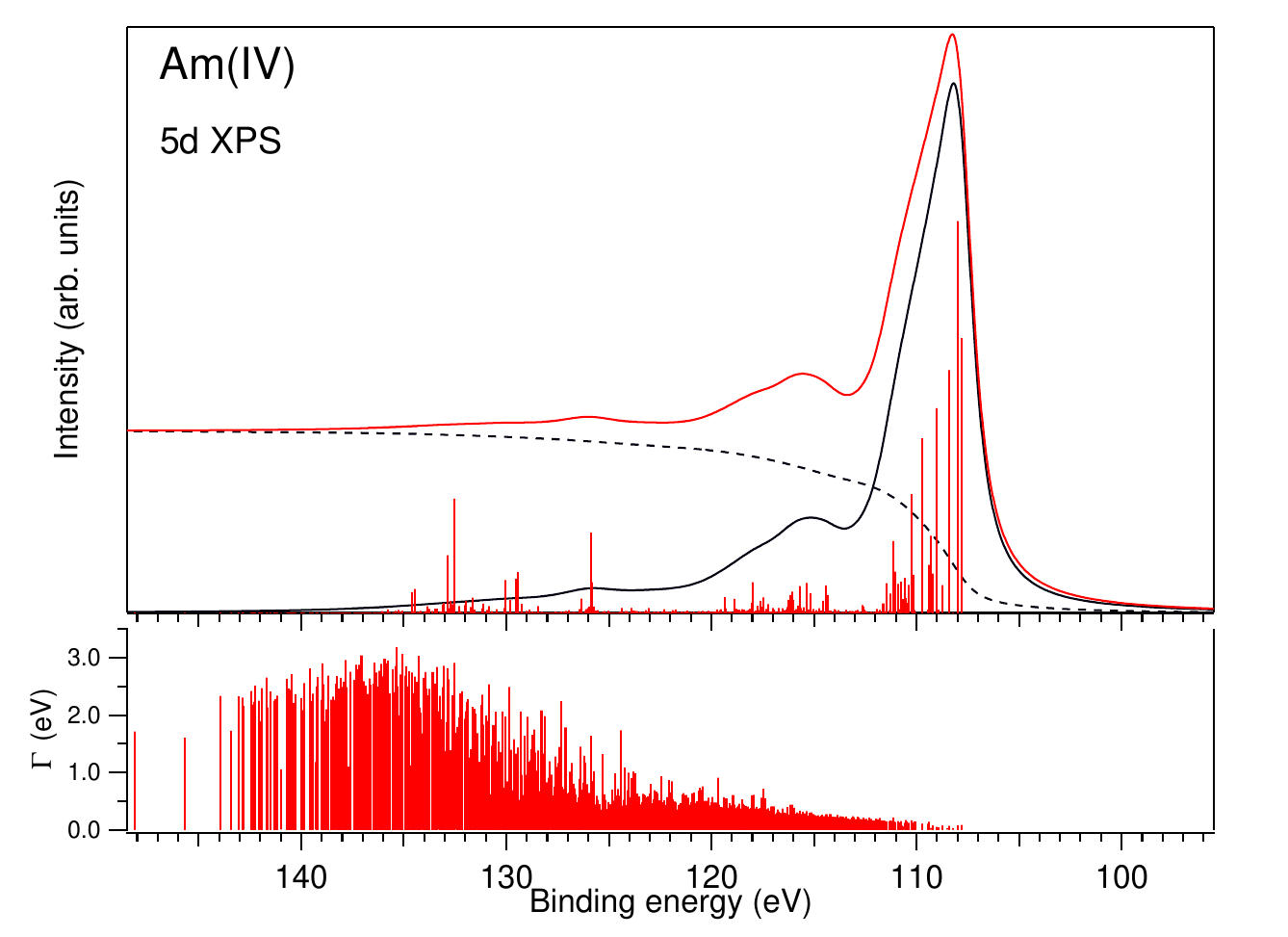}
\caption{Calculated $5d$ XPS spectrum of Am(IV) by taking into account the Auger decay of states of the $5d^95f^5\varepsilon{l}$ configuration and corresponding multiplet dependent lifetime broadening $\Gamma$. \label{AmO2_5dXPS}}
\end{figure}

\begin{figure}
\includegraphics[width=\columnwidth]{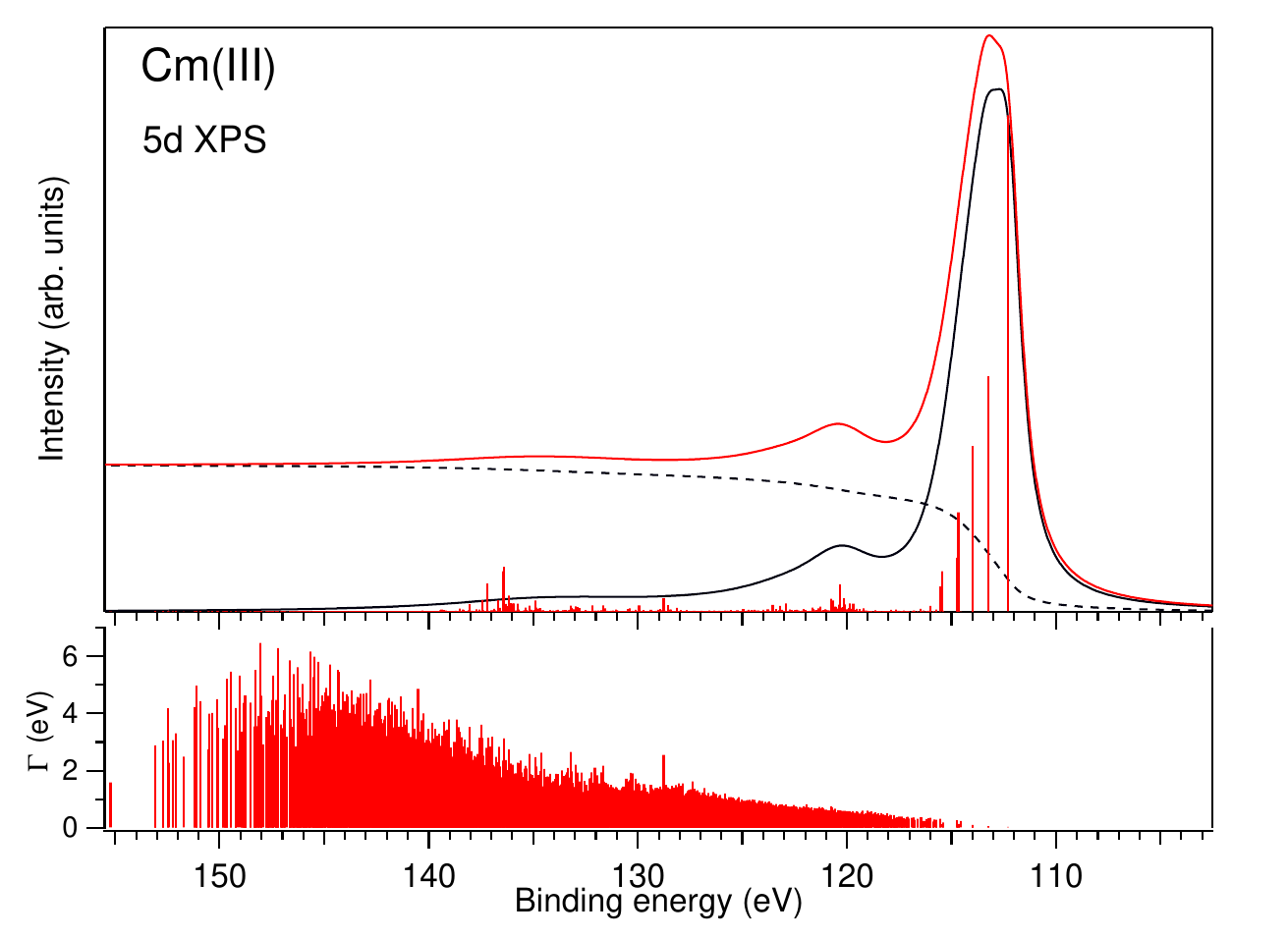}
\caption{Calculated $5d$ XPS spectrum of Cm(III) by taking into account the Auger decay of states of the $5d^95f^7\varepsilon{l}$ configuration and corresponding multiplet dependent lifetime broadening $\Gamma$. \label{Cm2O3_5dXPS}}
\end{figure}

The calculated $5d$ XPS spectra along with the corresponding $\Gamma$ values are shown in Figs.~\ref{UO2_5dXPS}$-$\ref{AmO2_5dXPS}. A comparison of the calculated $5d$ XPS spectrum of CmO$_2$ in Fig.~\ref{AnO2_Oh_5dXPS} with the experimental spectrum of Cm oxide recorded by Veal \textit{et al.} \cite{Veal} reveals noticeable differences. In particular, the width of the main $5d_{5/2}$ line of the measured spectrum appears to be much wider suggesting a contribution from several intense multiplet states instead of a single peak in the calculated spectrum (see Fig.~\ref{AnO2_Oh_5dXPS}). Furthermore, the $4f$ XPS spectrum recorded by Veal \textit{et al.} \cite{Veal} on the same oxide sample looks similar to that of Cm$_2$O$_3$ (Ref. \cite{Gouder}) and does not show the signs of expected charge-transfer satellites \cite{Yamazaki}. Based on results of the RIXS measurements \cite{Kvashnina} at the $5d$ edge of the Cm oxide, it was also discussed that the initial creation of Cm$_2$O$_3$ is more favorable upon Cm oxidation. Therefore, the $5d$ XPS spectrum of Cm(III) was also calculated and shown in Fig.~\ref{Cm2O3_5dXPS}. The Slater integrals values for the Cm(III) case are listed in Tables~\ref{table1},~\ref{table2} and \ref{table4} (see also Ref. \cite{Butorin_AcIII}).

\begin{table}
\caption{Hartree-Fock calculated values of Slater integrals for the $5d$-$5f5f(\varepsilon{g})$ super Coster-Kronig decay. They refer to the decay per square root of Rydberg. In the XPS calculations these values were reduced to 80\% after a conversion from Rydberg to eV units.}
\begin{tabular}{lccccc}
Parameter&U(IV)&Np(IV)&Pu(IV)&Am(IV)&Cm(III)\\
\hline
$R^1(5d\varepsilon{g},5f^2)$&4.711&4.828&4.780&4.878&4.774\\
$R^3(5d\varepsilon{g},5f^2)$&2.906&2.983&3.020&3.083&3.015\\
$R^5(5d\varepsilon{g},5f^2)$&2.043&2.100&2.147&2.193&2.143\\
\end{tabular}
\label{table4}
\end{table}

The $5d$-$5f5f(\varepsilon{g})$ super Coster-Kronig decay does not lead to a significant smearing of the calculated U(IV) $5d$ XPS peak at the binding energy of $\sim$118.4 eV (Fig.~\ref{UO2_5dXPS}). Similar situation occurs for the  Np(IV) $5d$ XPS peak at the binding energy of $\sim$122.1 eV (Fig.~\ref{NpO2_5dXPS}). Therefore, in addition, the $5d$-$5f6p(\varepsilon{l})$ Coster-Kronig decay was also taken into account in the $5d$ XPS calculations for U(IV) and Np(IV). That is not surprising because the non-negligible probability for such a Coster-Kronig decay was shown for the $4d$ XAS of light lanthanides \cite{Ogasawara_4dXAS}. The final results are represented in Figs.~\ref{UO2_5dXPS} and \ref{NpO2_5dXPS}.

An inspection of Figs.~\ref{UO2_5dXPS}$-$\ref{Cm2O3_5dXPS} shows that in systems with larger number of $5f$ electrons, such as AmO$_2$ and Cm$_2$O$_3$, the corresponding multiplet dependent lifetime broadening $\Gamma$ grows gradually towards to higher binding energies, while for lighter actinides, $\Gamma$ exhibits a step-like increase. In either case, the $5d$ XPS transitions at high binding energies become strongly smeared out.

In the $5d$ XPS spectrum of ThO$_2$, the satellite at the binding energy of $\sim$7.5 eV above the $5d_{3/2}$ line is observed \cite{Veal,Putkov} which was suggested to have the Th $5f-$O $2p$ charge-transfer character in analogy to that in the $4f$ XPS spectrum of ThO$_2$ (Ref. \cite{Yamazaki}). Therefore, the AIM calculation of the $5d$ XPS spectrum of ThO$_2$ was performed to take into account the Th $5f-$O $2p$ hybridization. The ground state of ThO$_2$ was described as a linear combination of the $5f^{0}$,  $5f^{1}\underline{\upsilon}^{1}$ and $5f^{2}\underline{\upsilon}^{2}$ configurations, where $\underline{\upsilon}$ stands for an electronic hole in the O $2p$ band. The final state of the XPS process was represented by a combination of the $5d^{9}5f^{0}$, $5d^{9}5f^{1}\underline{\upsilon}^{1}$ and $5d^{9}5f^{2}\underline{\upsilon}^{2}$ configurations. In the limit of $V\rightarrow0$ (see the Computational details section), the difference between the configuration averaged energies for the ground state can be written as $E(5f^{n+1}\underline{\upsilon}^{1})-E(5f^{n})=\Delta$ and $E(5f^{n+2}\underline{\upsilon}^{2})-E(5f^{n+1}\underline{\upsilon}^{1}))=\Delta+U_{ff}$, where $\Delta$ is the actinide $5f-$O $2p$ charge-transfer energy and $n$ is equal 0 in the ThO$_2$ case. $\Delta$ is taken as $\Delta=\varepsilon_{5f}-\varepsilon_{\upsilon}^0$. For the final state of the XPS process, the difference between the configuration averaged energies can be defined as $E(5d^{9}5f^{n+1}\underline{\upsilon}^{1})-E(5d^{9}5f^{n})=\Delta-U_{fc}$ and $E(5d^{9}5f^{n+2}\underline{\upsilon}^{2})-E(5d^{9}5f^{n+1}\underline{\upsilon}^{1})=\Delta+U_{ff}-U_{fc}$.

\begin{figure}
\includegraphics[width=\columnwidth]{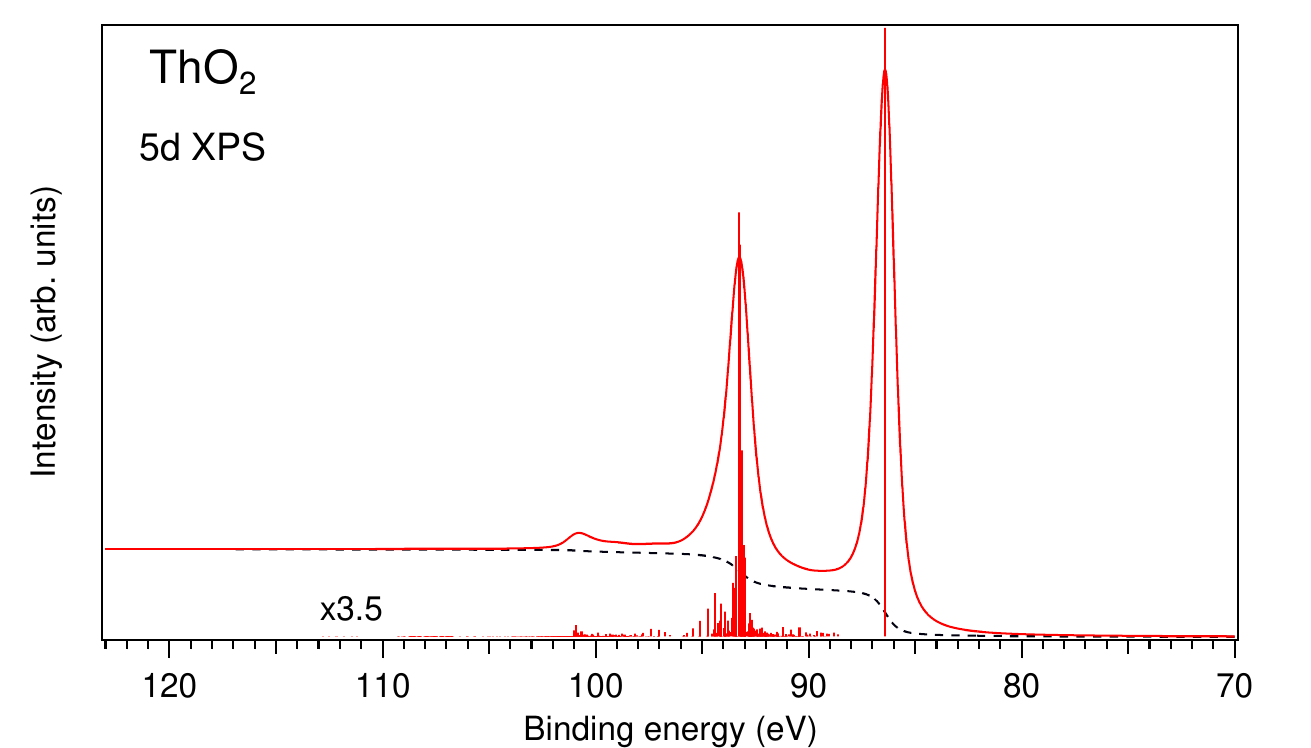}
\caption{Calculated $5d$ XPS spectrum of ThO$_2$. For clarity, the intensity of the multiplet poles is multiplied by 3.5 so that the main pole at 86.4 eV is cut by the upper border of the graph frame at 22\% of its intensity/height. \label{ThO2_5dXPS}}
\end{figure}

To reproduce the experimental $5d$ XPS spectrum of ThO$_2$, the following values of the model parameters were used in the AIM calculations: $\Delta$=8.0 eV, $U_{ff}$=4.0 eV, $U_{fc}$=4.5 eV, $V$=1.2 eV. The values are similar to those used for the calculation of Th $M_{4}$ HERFD-XAS spectrum of ThO$_2$ (Ref. \cite{Butorin_PNAS}) except for the $U_{fc}$ value, because it is expected to be smaller for the shallow $5d$ core hole. For the final state of the $5d$ XPS process, $V$ was reduced to 1.05 eV since the degree of the $4f$ localization increases due to the interaction with the core hole \cite{Gunnarsson}. The $N$ parameter was set to 8 and $W$ was taken to be 2.0 eV. The $F^{k}(5f,5f)$, $F^{k}(5d,5f)$ and $G^{k}(5d,5f)$ integrals were reduced to 80\%, 75\% and 65\%, respectively, of their \textit{ab-initio} Hartree-Fock values determined for all electronic configurations included in the AIM calculation of the $5d$ XPS spectrum of ThO$_2$. Note, that the effect of the Th $6d-$O $2p$ hybridization (and $6d$-$5f$ interaction) in the spectrum \cite{Butorin_PNAS} was disregarded for simplicity, because the corresponding satellite is expected to be hidden under the multiplet structure around the $5d_{3/2}$ line as well as smeared out due to the $5d_{3/2}$ broadening and because the interaction of the $5d$ core hole with $5f$ electrons is significantly stronger than that for the $3d$ core hole, as in case of Th $M_{4}$ HERFD-XAS \cite{Butorin_PNAS}. Fig.~\ref{ThO2_5dXPS} displays the AIM-calculated $5d$ XPS spectrum of ThO$_2$ both without and with the added background. The charge-transfer satellite is located at the energy of $\sim$7.4 eV above the $5d_{3/2}$ line and very weak transitions exist in the region between 13.5 eV and 16.5 eV above the $5d_{3/2}$ line (within the 106.5-109.5 eV binding energy range).

\begin{figure*}
\includegraphics[width=0.8\textwidth]{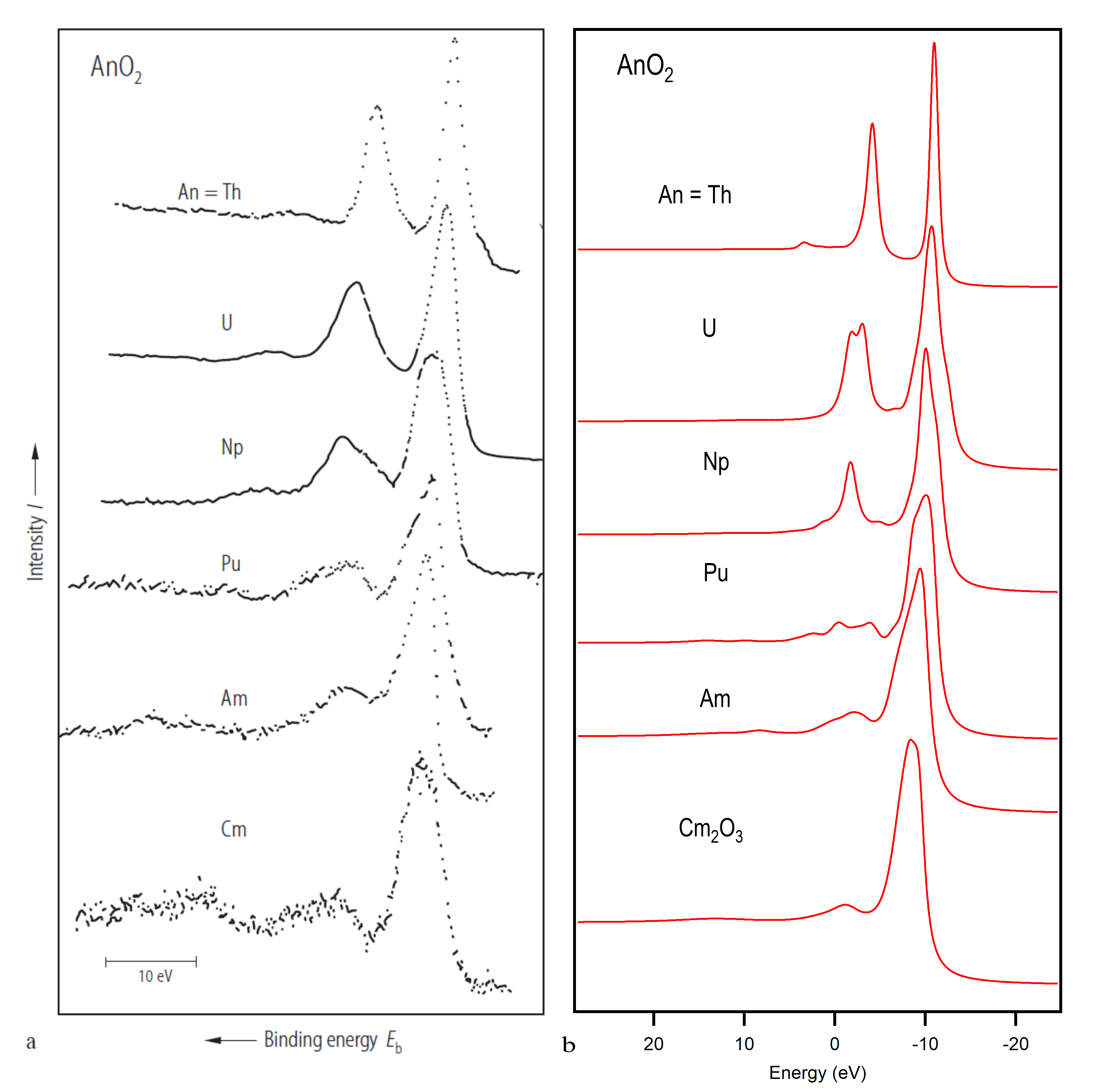}
\caption{Experimental (\textbf{a}) and calculated (\textbf{b}) $5d$ XPS spectra of several actinide oxides. The calculated spectra (on the right side) are arbitrarily aligned for a better comparison with experimental ones (on the left side). The experimental data were published by Veal \textit{et al.} \cite{Veal}. \label{AnO2_5dXPS}}
\end{figure*}

Fig.~\ref{AnO2_5dXPS} presents a comparison of the calculated $5d$ XPS spectra with the experimental ones by Veal \textit{et al.} \cite{Veal} for several actinide oxides.  Although, there are more recent experimental data published separately for some of these oxides \cite{Beatham,Ilton2,Allen,Teterin_PuO2,Teterin_NpO2,Teterin_AmO2,Putkov}, the data by Veal \textit{et al.} \cite{Veal} were measured for a whole series. An inspection of Fig.~\ref{AnO2_5dXPS} indicates an overall good agrement between the calculations and experiment. However, some differences with respect to the spectral shape are encountered, in particular for NpO$_2$. The structure in the $5d_{3/2}$ region appears to be substantially wider (see also \cite{Teterin_NpO2}) than that in the calculated spectrum. NpO$_2$ has a complex magnetic structure in the ground state \cite{Suzuki} which was not taken into account in our calculations. Some structures, which are observed at high binding energies in the spectra of PuO$_2$, AmO$_2$ and Cm oxide in Fig.~\ref{AnO2_5dXPS}a, were not observed in more recent data, at least for PuO$_2$ and AmO$_2$ (Refs. \cite{Teterin_PuO2,Teterin_AmO2}). The calculated spectra indicate a substantial smearing of the high binding energy structures due to the Auger decay (Fig.~\ref{AnO2_5dXPS}b). However, the Auger decay rate may be somewhat overestimated in atomic calculations when compared to solids.

The results of the calculations for Cm(III) describe better the main structures of the experimental spectrum of Cm oxide in Fig.~\ref{AnO2_5dXPS}a than those for Cm(IV), thus suggesting that the Cm oxide sample in Ref. \cite{Veal} had a composition close to Cm$_2$O$_3$. The curium case indicates that the $5d$ XPS spectra can be quite useful for distinguishing between oxidation states. As a whole, the significant degree of the $5f$ localization and rich, extended multiplet structure due to the strong interaction of the $5d$ core hole with $5f$ electrons in actinide oxides (except for Th) allows one to reproduce the $5d$ XPS spectra with help of the atomic/crystal-field multiplet approach and use these spectra for the characterization of the actinide oxidation state.

\subsection{Lanthanide $3d$ XPS}

As in the ThO$_2$ case, the lanthanide $4f-$O $2p$ hybridization and charge-transfer effects were discussed for the description of the $3d$ XPS spectra of Ce, Pr and Tb dioxides \cite{Kotani,Nakazawa,Kolorenc,Ikeda}. However, the AIM calculations \cite{Kotani,Ikeda} for PrO$_2$ and TbO$_2$ were simplified and did not include the full multiplet effects and in particular the spin-orbit interaction for the $3d$ level.

\begin{figure}
\includegraphics[width=\columnwidth]{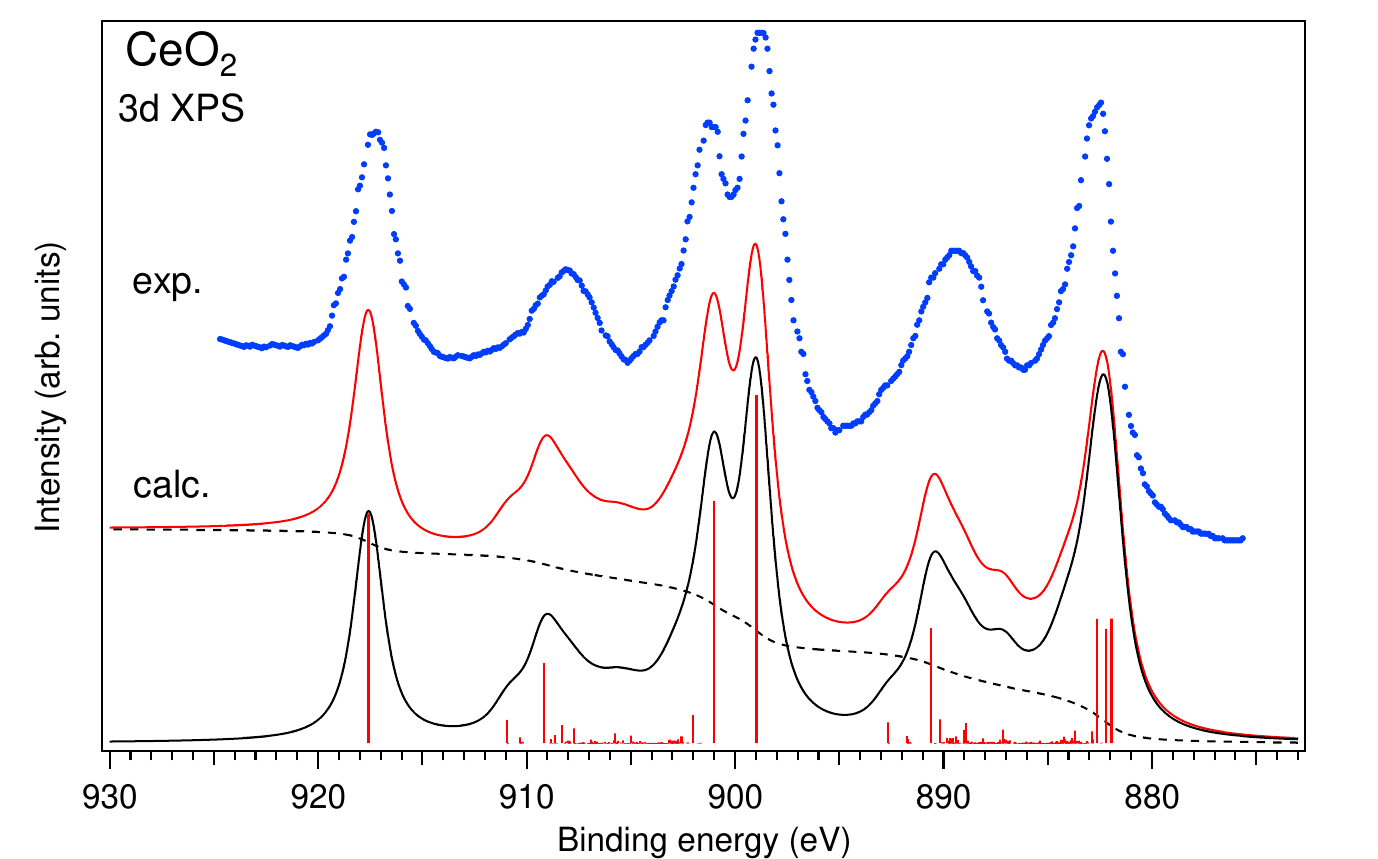}
\caption{Experimental and AIM-calculated $3d$ XPS spectra of CeO$_2$. Calculated results: the solid black curve is the convoluted spectrum, the dashed black curve represents the background and the solid red curve is the sum of the convoluted spectrum and background. \label{CeO2_3dXPS}}
\end{figure}

\begin{figure}
\includegraphics[width=\columnwidth]{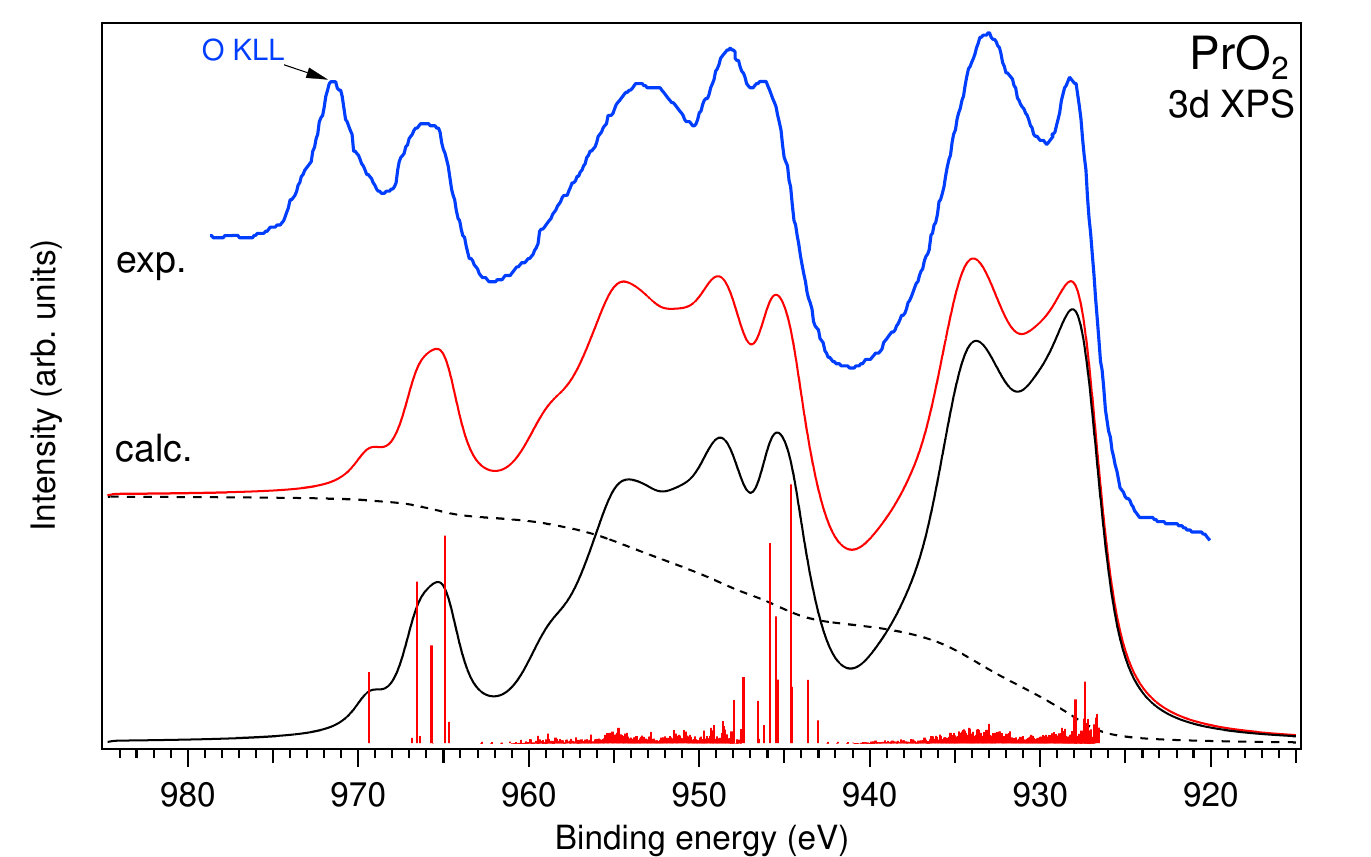}
\caption{Experimental and AIM-calculated $3d$ XPS spectra of PrO$_2$. The experimental spectrum was adopted from Ref. \cite{Schaefer}. \label{PrO2_3dXPS}}
\end{figure}

\begin{figure}
\includegraphics[width=\columnwidth]{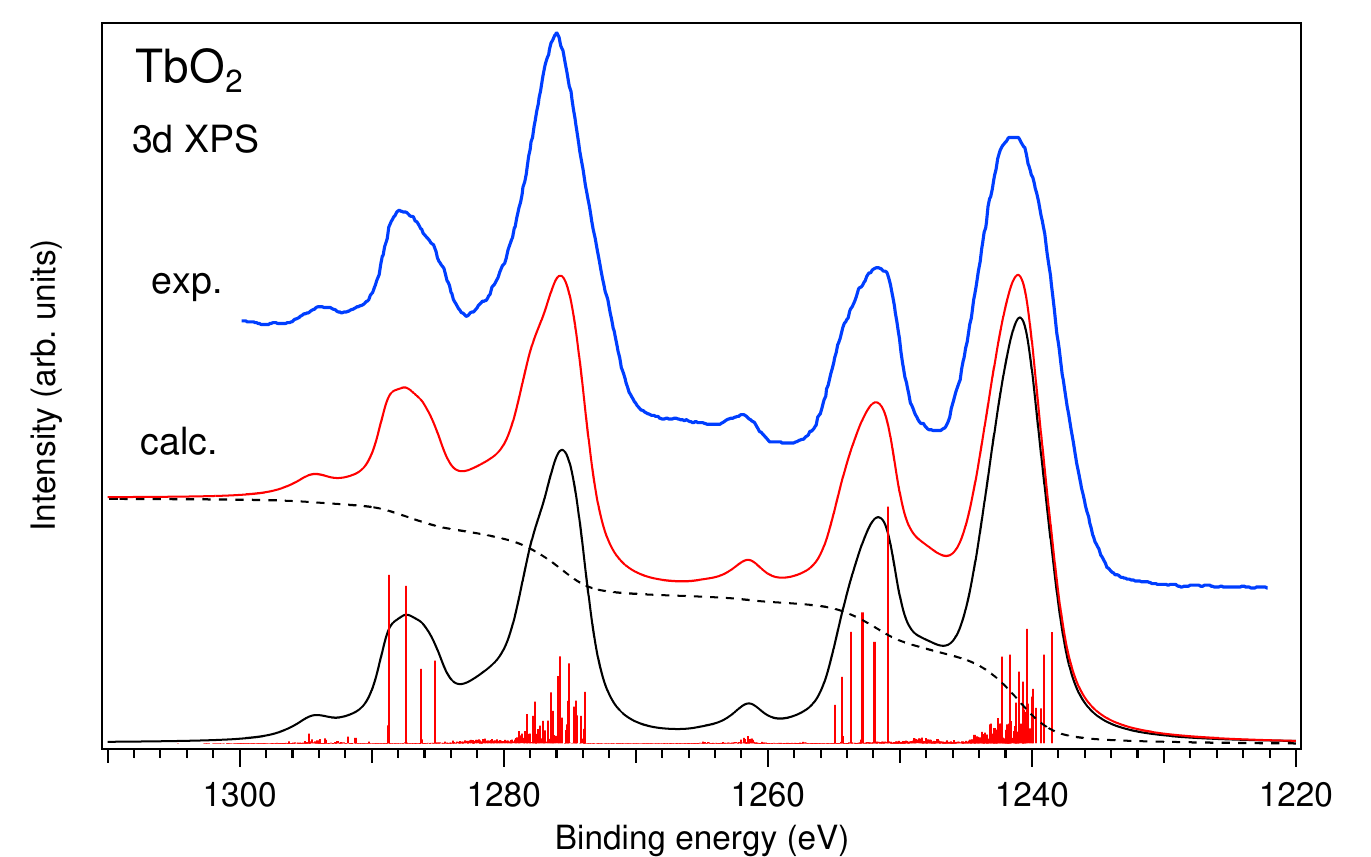}
\caption{Experimental and AIM-calculated $3d$ XPS spectra of TbO$_2$. The experimental spectrum was adopted from Ref. \cite{Blanco}.\label{TbO2_3dXPS}}
\end{figure}

Figs.~\ref{CeO2_3dXPS}-\ref{TbO2_3dXPS} show a comparison of experimental $3d$ XPS spectra with AIM-calculated ones in this work for CeO$_2$, PrO$_2$ and TbO$_2$, respectively. The values of the Slater integrals defining the multiplet structure in the XPS calculations are listed in Table~\ref{table5}. The ground state of the XPS process was described by a mixture of the $5f^{n}$,  $5f^{n+1}\underline{\upsilon}^{1}$ and $5f^{n+2}\underline{\upsilon}^{2}$ configurations and the final state was described by a combination of the $3d^{9}5f^{n}$, $3d^{9}5f^{n+1}\underline{\upsilon}^{1}$ and $3d^{9}5f^{n+2}\underline{\upsilon}^{2}$ configurations, where $n$=0, 1, 7, respectively. The values of the AIM parameters used in the $3d$ XPS calculations for CeO$_2$, PrO$_2$ and TbO$_2$ are listed in Table~\ref{table6}. For CeO$_2$ and PrO$_2$, the value of lanthanide $4f-$O $2p$ hopping term $V$ was determined from the analysis of $3d$-$4f$ RIXS \cite{Butorin_CeO2,Butorin_PrO2} which probes the electronic structure in the ground state. To take into account the configuration dependence of $V$ (Ref. \cite{Gunnarsson}), it was multiplied by reduction factor $\kappa$ (see Table~\ref{table6}) for the final state of the $3d$ XPS process to get a better fit of the $3d$ XAS \cite{Butorin_CeO2,Butorin_PrO2} and $3d$ XPS spectra. The applied core-hole lifetime broadening is indicated in Table~\ref{table6} and the experimental energy resolution was simulated by the Gaussian with HWHM of 0.3 eV.

\begin{table*}
\caption{\textit{Ab-initio} Hartree-Fock values (in units of eV) of Slater integrals and spin-orbit coupling constants for Ce, Pr and Tb. In the XPS calculation, the $F^{k}(4f,4f)$, $F^{k}(3d,4f)$ and $G^{k}(3d,4f)$ integrals were reduced to 80\% of these values.}
\begin{tabular}{lcccccccccc}
Configuration&$F^{2}(4f,4f)$&$F^{4}(4f,4f)$&$F^{6}(4f,4f)$&$\zeta(4f)$&$F^{2}(3d,4f)$&$F^{4}(3d,4f)$&$G^{1}(3d,4f)$&$G^{3}(3d,4f)$&$G^{5}(3d,4f)$&$\zeta(3d)$\\
\hline
Ce $4f^0$& & & & & & & & & & \\
Ce $4f^1$& & & &0.087& & & & & & \\
Ce $4f^2$&10.223&6.347&4.548&0.074& & & & & & \\
Ce $3d^94f^0$& & & & & & & & & &7.439\\
Ce $3d^94f^1$& & & &0.119&8.194&3.751&5.658&3.312&2.287&7.442\\
Ce $3d^94f^2$&12.628&7.940&5.717&0.107&7.486&3.384&5.073&2.968&2.048&7.446\\
Pr $4f^1$& & & &0.115& & & & & & \\
Pr $4f^2$&12.227&7.670&5.518&0.102& & & & & & \\
Pr $4f^3$&10.886&6.770&4.854&0.089& & & & & & \\
Pr $3d^94f^1$& & & &0.148&9.221&4.293&6.538&3.832&2.647&8.131\\
Pr $3d^94f^2$&14.100&8.918&6.436&0.136&8.568&3.944&5.974&3.499&2.416&8.135\\
Pr $3d^94f^3$&13.096&8.235&5.929&0.123&7.889&3.591&5.410&3.166&2.185&8.139\\
Tb $4f^7$&15.829&9.981&7.195&0.237& & & & & & \\
Tb $4f^8$&14.915&9.360&6.734&0.221& & & & & & \\
Tb $4f^9$&13.892&8.670&6.225&0.205& & & & & & \\
Tb $3d^94f^7$&17.274&10.946&7.907&0.285&11.199&5.319&8.229&4.830&3.338&13.358\\
Tb $3d^94f^8$&16.461&10.390&7.493&0.268&10.631&5.013&7.730&4.535&3.133&13.363\\
Tb $3d^94f^9$&15.586&9.794&7.050&0.251&10.055&4.709&7.240&4.245&2.933&13.368\\
\end{tabular}
\label{table5}
\end{table*}

\begin{table}
\caption{AIM parameter values (in units of eV) used in calulations of $3d$ XPS spectra of lanthanide oxides. $\kappa$ is the reduction factor/coefficient for $V$ in the final state of the XPS process and $n_f$ is the calculated $4f$-shell occupancy (number of electrons) in the ground state.}
\begin{tabular}{lccc}
Parameter&CeO$_2$&PrO$_2$&TbO$_2$\\
\hline
$\Delta$&2.0&0.5&-4.0\\
$U_{ff}$&9.0&8.0&8.0\\
$U_{fc}$&12.6&11.0&10.5\\
$V$&1.0&0.8&0.8\\
$\kappa$&0.7&0.7&0.8\\
$\Gamma$&0.8&1.0&1.0\\
$n_f$&0.54&1.72&7.53\\
\end{tabular}
\label{table6}
\end{table}

The experimental $3d$ XPS spectrum of single-crystal CeO$_2$ in Fig.~\ref{CeO2_3dXPS} was measured using the monochromatic Al $K\alpha$ radiation and its analysis in terms of the AIM calculations indicated that the peaks at binding energies of $\sim$898.8 eV and $\sim$917.3 eV have almost pure $3d^94f^0$ character while other structures are mainly a mixture of $3d^{9}4f^{1}\underline{\upsilon}^{1}$ and $3d^{9}4f^{2}\underline{\upsilon}^{2}$ configurations. Nakazawa \textit{et al.} \cite{Nakazawa} have already calculated the $3d$ XPS spectrum of CeO$_2$ using the AIM-parameters derived from the analysis of $3d$-$4f$ RIXS of CeO$_2$ (Ref. \cite{Butorin_CeO2}). Here, we performed the calculations with almost the same AIM parameters (reduction factor $\kappa$ for $V$ in a core-hole state was 0.7 instead of 0.6) in order to have a complete set of the calculated spectra for lanthanide dioxides.

For PrO$_2$, the $3d$ XPS calculations were also carried out with the AIM parameters which were chosen based on the analysis of $3d$-$4f$ RIXS and $3d$ XAS of this oxide \cite{Butorin_PrO2}. The contribution of different electronic configurations to the spectral structures are similar to the case of CeO$_2$. The peaks at binding energies of $\sim$946.0 eV and $\sim$965.9 eV have mainly the $3d^94f^1$ origin and other structures consist of a mixture of the $3d^{9}4f^{2}\underline{\upsilon}^{1}$ and $3d^{9}4f^{3}\underline{\upsilon}^{2}$ configurations (see Fig.~\ref{PrO2_3dXPS}).

For TbO$_2$, the $3d$-$4f$ RIXS measurements are not available, but the AIM parameter values used in the $3d$ XPS calculations were found to be close to those determined for BaTbO$_3$ (Ref. \cite{Hu}). The structures at binding energies of $\sim$1251.5 eV and $\sim$1287.2 eV originate mainly from the $3d^94f^7$ configuration while the main spectral lines are contributed by configurations with the larger number of the $4f$ electrons (see Fig.~\ref{TbO2_3dXPS}). The small satellites at binding energies of $\sim$1261.6 eV and $\sim$1294.0 eV also belong to the $3d^94f^7$ configuration and were attributed to the spin-flip states in Ref. \cite{Hu}. A negative value for $\Delta$ comes as a result of its definition as the energy difference between the average of the two configurations which have an extended multiplet structure in this oxide while the ground state is mainly defined by a mixture of the lowest states of those configurations.

As a whole, an increase in occupancy $n_f$ due to the $f-$O $2p$ hybridization and covalency of the chemical bonding derived from the analysis of the XPS spectra seems to be higher in lanthanide dioxides as compared to actinide dioxides \cite{Yamazaki}. It would be interesting to investigate what happen to the chemical bonding when Pr and Tb enter to the UO$_2$ and PuO$_2$ lattice (e.g. as fission products) without a creation of clusters (see e.g. Ref. \cite{Bazarkina}).

Kotani and Ogasawara \cite{Kotani} reported a trend in changes of the AIM parameter values for lanthanide sesquioxide series throughout the whole lanthanide row based on the calculations of the $3d$ XPS spectra. However, the analysis of the data for Tb(III) was omitted. Therefore, we added here the description of the $3d$ XPS spectrum of Tb$_2$O$_3$.

\begin{figure}
\includegraphics[width=\columnwidth]{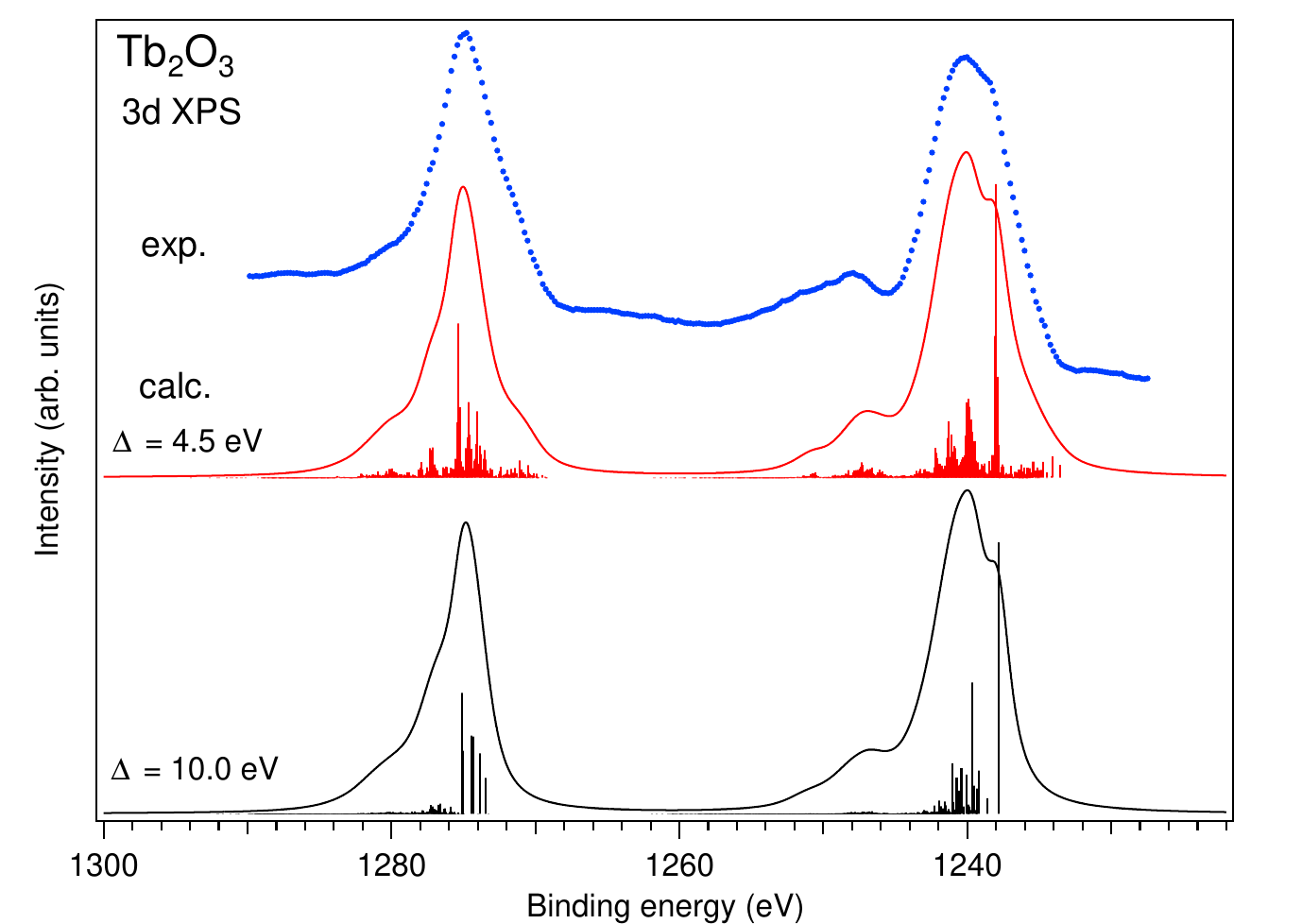}
\caption{Comparison of the experimental $3d$ XPS spectrum of Tb$_2$O$_3$ with ones calculated using AIM with different $\Delta$ values. \label{Tb2O3_3dXPS}}
\end{figure}

Fig.~\ref{Tb2O3_3dXPS} compares the experimental data with results of calculations. The experimental spectrum of the Tb$_2$O$_3$ film was recorded using the monochromatic Al $K\alpha$ radiation. In calculations, the ground state was represented by a linear combination of the $4f^8$ and $4f^{9}\underline{\upsilon}^{1}$ configurations because the contribution of the $4f^{9}\underline{\upsilon}^{2}$ is expected to be very small. The final state of the spectroscopic process was described by a combination of the $3d^94f^8$ and of $3d^{9}4f^{9}\underline{\upsilon}^{1}$ configurations. By interpolating the data from Ref. \cite{Kotani}, the following AIM parameter values were used: $\Delta$=10.0 eV, $U_{ff}$=9.5 eV, $U_{fc}$=10.5 eV, $V$=0.45 eV. In the final state of the $3d$ XPS process, the hybridization strength was reduced to $V$=0.36 eV ($\kappa$=0.8). While the AIM-calculated spectrum using these parameter values (see the lowest curve in Fig.~\ref{Tb2O3_3dXPS}) describes most of the structures in the experimental spectrum, the asymmetry in the shape on the low binding-energy sides of the main lines is not reproduced.

An inspection of the differences in the $\Delta$ value between CeO$_2$ and Ce$_2$O$_3$ and between PrO$_2$ and Pr$_2$O$_3$ (Ref. \cite{Kotani}) suggests that the the $\Delta$ in Tb$_2$O$_3$ is probably lower than 10.0 eV. Indeed, the AIM calculations of the $3d$ XPS spectrum with $\Delta$=4.5 eV allow one to reproduce quite well the experimental spectrum of Tb$_2$O$_3$ (see Fig.~\ref{Tb2O3_3dXPS}). This result is not consistent with the trend in the $\Delta$ behavior for lanthanide sesquioxides as described by Kotani and Ogasawara \cite{Kotani} and suggests more complex changes around the middle of the lanthanide row. More measurements with other spectroscopic methods, in particular the high-resolution $3d$-$4f$ RIXS data, may help to clarify this issue.

\section{Conclusion}
Regarding, in general, the higher degree of the localization of the $4f$ shell compared to the $5f$ shell, the higher covalency of the chemical bonding in lanthanide dioxides compared to that in actinide dioxides, which is revealed from the analysis of the XPS data, is somewhat unusual to expect. However, only very few lanthanide dioxides exist in a stable form, thus emphasizing the uniqueness of the situation.

\section{Acknowledgement}
The research was funded by the European Union's "EURATOM" research and innovation program under grant agreement No. 101164053.

\bibliography{An_5dXPS}

\end{document}